\documentstyle[preprint,epsf,eqsecnum,aps]{revtex}
\input{epsf.sty}
\tightenlines

\begin{document}
\draft
\title{Low-temperature properties of the spin-1 antiferromagnetic Heisenberg chain with bond-alternation
}
\author{Masanori Kohno$^{1}$, Minoru Takahashi$^{1}$, and Masayuki Hagiwara$^{2}$}
\address{
$^1$Institute for Solid State Physics,\\
University of Tokyo, Roppongi, Minato-ku, Tokyo 106, Japan\\
$^2$The Institute of Physical and Chemical Research (RIKEN), Wako, 
Saitama 351-01, Japan}
\date{\today}
\maketitle
\begin{abstract}
We investigate the low-temperature properties of the spin-1 antiferromagnetic Heisenberg chain with bond-alternation by the quantum Monte Carlo method (loop algorithm). The strength of bond-alternation at the gapless point is estimated as $\delta_{\rm c}=0.2595\pm0.0005$. We confirm numerically that the low-temperature properties at the gapless point are consistent with field theoretical predictions. 
The numerical results are compared with those of the spin-1/2 antiferromagnetic Heisenberg chain 
and recent experimental results for [\{Ni(333-tet)($\mu$-N$_3$)\}$_n$](ClO$_4$)$_n$ (333-tet=tetraamine 
$N,N^{\prime}$-bis(3-aminopropyl)-1,3-propanediamine).
\end{abstract}

\narrowtext

\section{Introduction}
\label{Introduction}

For one-dimensional quantum spin chains, there are some theoretical predictions which have been confirmed experimentally.
One example is the Haldane conjecture\cite{Haldane}. 
Haldane investigated the O(3) nonlinear $\sigma$ model, and predicted that an excitation gap opens for integer spin chains but not for half-odd integer spin chains. This conjecture was confirmed by numerical calculations\cite{Botet,ParkinsonBonner,Nightingale}, and the Haldane gap was observed experimentally\cite{Buyers,Renard,Katsumata}.
Another example of such theoretical predictions is the presence of logarithmic corrections at very low temperatures for the spin-1/2 antiferromagnetic Heisenberg chain (S1/2AH).
This feature was predicted by the renormalization-group approach\cite{AGSZ,Nomura} and was confirmed numerically by Bethe Ansatz\cite{EAT}. Later, this behavior was observed experimentally\cite{S1/2AHexp}. In this way, the presence of logarithmic corrections at very low temperatures for the spin-1/2 antiferromagnetic Heisenberg chain was established.
\par
There is another interesting prediction which has been observed experimentally quite recently\cite{HagiwaraUs}. This is the so-called Affleck-Haldane conjecture\cite{Affleck,AffleckHaldane}. Affleck extended Haldane's argument to the bond-alternating chains, and predicted that there will be $2S$ gapless points for the spin-$S$ antiferromagnetic Heisenberg chain with bond-alternation. Later, for the spin-1 case, the central charge $c$ at the gapless point is estimated as $c\simeq 1$ by numerical calculations\cite{KatoTanaka,Yamamoto1,Totsuka}. This implies that the low-lying excitations at the gapless point are described by the level $k=1$ SU(2) Wess-Zumino-Witten (WZW) model. As a result, the same qualitative low-temperature properties as those of the spin-1/2 antiferromagnetic Heisenberg chain are expected at the gapless point.
Recent experiments for [\{Ni(333-tet)($\mu$-N$_3$)\}$_n$](ClO$_4$)$_n$ (333-tet=tetraamine 
$N,N^{\prime}$-bis(3-aminopropyl)-1,3-propanediamine) show that this compound has a structure which is effectively described by the spin-1 antiferromagnetic Heisenberg chain with bond-alternation (S1BA)\cite{HagiwaraSt}, and that the behavior of the uniform susceptibility is close to what is expected at the gapless point\cite{HagiwaraUs}. Thus, this compound is probably an experimental realization of the gapless point of the Affleck-Haldane conjecture.
In order to clarify this feature in more detail, it is necessary to show explicitly how the uniform susceptibility behaves in the low-temperature regime for S1BA at the gapless point. Hence, the purpose of this paper is to clarify the range of temperatures for which the field-theoretical prediction is valid, and to what extent S1BA can explain the low-temperature properties of this compound.
\par
For the gapless point of S1BA, Singh and Gelfand applied a series expansion technique, and estimated the critical value of the strength of bond-alternation $\delta$ as $\delta_{\rm c}=0.25\pm0.03$\cite{Singh}. Later, Kato and Tanaka applied the density-matrix renormalization group (DMRG) method, and obtained clear evidence of the Affleck-Haldane conjecture\cite{KatoTanaka}. They estimated the critical value $\delta_{\rm c}$ more accurately as $\delta_{\rm c}=0.25\pm0.01$. As for the excitation spectrum, Yamamoto performed the quantum Monte Carlo simulation (world-line algorithm), and obtained the dispersion relation\cite{Yamamoto1}. Totsuka, {\it et.al.} investigated the low-lying excitation spectrum by exact diagonalization, and analyzed the results using the conformal field theory\cite{Totsuka}. They estimated the critical value of $\delta$ as $\delta_{\rm c}=0.254\pm0.008$. 
\par
Before investigating the low-temperature properties at the gapless point, 
we have to determine the critical value of $\delta$ more accurately. 
In Sec.\ref{Gapless}, we estimate $\delta_{\rm c}$ from the low-temperature behavior of the uniform susceptibility and the staggered susceptibility.
In Sec.\ref{Main}, we analyze the low-temperature properties of S1BA at the gapless point based on the field-theoretical prediction. 
Comparisons with numerical results of the spin-1/2 antiferromagnetic Heisenberg chain are also made. 
In Sec.\ref{Experiment}, the numerical results are compared with recent experimental results 
for [\{Ni(333-tet)($\mu$-N$_3$)\}$_n$](ClO$_4$)$_n$. 
\par
In the present paper, we consider S1BA defined by the following Hamiltonian:
\begin{equation}
   {\cal H} = J\sum_i[1-(-1)^i\delta]
                    \mbox{\boldmath $S_i\cdot S_{i+1}$},
\end{equation}
where $\mbox{\boldmath $S_i$}$ denotes the spin operator at site $i$ 
with spin one ($S=1$).
The length of the chain and the temperature are denoted by $L$ and $T$, respectively. We set $J=1$ as the energy unit.
\par
We use the quantum Monte Carlo (QMC) method (loop algorithm)\cite{loop_alg}. The simulations have been performed in the grand-canonical ensemble. We have typically run $6\times10^5$ Monte Carlo steps for measurements after $6\times10^4$ steps. We have made extrapolation for the Trotter slice $\Delta\tau$ ($\Delta\tau\rightarrow0$) as $a+b\cdot\Delta\tau^2$ using the data at $\Delta\tau\simeq 1/6,1/7$ and $1/8$. The system-sizes we have investigated are $L=32,64,96,192,320$ and $400$.
We have performed calculations up to inverse temperature $\beta=100$. 

\section{Finite-temperature properties of S1BA}
\label{FiniteT}
First, let us look over global features of S1BA for finite temperatures.
Figure \ref{UsSsS0gl} shows the uniform susceptibility $\chi(q=0;T)$, the staggered susceptibility $\chi(q=\pi;T)$ and the staggered structure factor $S(q=\pi;T)$ defined as
\begin{eqnarray}
\chi(q=0;T) &\equiv& \frac{1}{TL}\langle(\sum_{i}S^z_i)^2\rangle_T,\\
\chi(q=\pi;T) &\equiv& \frac{1}{TL}\langle(\sum_{i,j}(-1)^iS^z_{(i,j)}/M)^2\rangle_T,\\
S(q=\pi;T) &\equiv& \frac{1}{L}\langle(\sum_{i}(-1)^iS^z_i)^2\rangle_T,
\end{eqnarray}
where $S^z_{(i,j)}$ denotes the $z$-component of the spin at site $(i,j)$ in the (1+1)-dimensional space-time, and $M$ is the number of Trotter slices defined as $M\equiv\beta/\Delta\tau$. Here, $\langle\cdots\rangle_T$ denotes the thermal average at temperature $T$.
\par
The uniform susceptibility $\chi(q=0;T)$ in the low-temperature regime is fitted well by $\chi(q=0;T)\propto\frac{1}{\sqrt{T}}\exp(-\Delta/T)$, except for $\delta=0.25$, as shown in Fig.\ref{Usglbeta}. This behavior is expected when the dispersion relation is $E(q)|_{q\rightarrow 0}=aq^2+\Delta$\cite{Troyer}, where $q$ is the momentum measured from the lowest triplet state and $a$ is a constant. Thus, except for $\delta=0.25$, the low-energy excitation may be explained by a magnon excitation with a finite excitation gap $\Delta$\cite{ParkinsonBonner,Takahashi}. This excitation gap $\Delta$ gradually decreases as $\delta$ approaches $\delta_{\rm c}\simeq 0.25$.
As for the staggered susceptibility $\chi(q=\pi;T)$ and the staggered structure factor $S(q=\pi;T)$, the maximum value becomes larger as $\delta$ approaches $\delta_{\rm c}\simeq 0.25$. This behavior is expected from the Affleck-Haldane conjecture and is consistent with various numerical results\cite{KatoTanaka,Yamamoto1,Totsuka,Yamamoto2}.

\section{Estimation of the critical point}
\label{Gapless}
In this section, we accurately determine the critical value $\delta_{\rm c}$. One criterion is based on the expectation that the uniform susceptibility $\chi(q=0;T)$ reaches a maximum at the critical point if the temperature is low enough.
In Fig.\ref{UsSsdelta}(a), we show the $\delta$-dependence of the uniform susceptibility.
The maximum is located near $\delta\simeq0.2595$ with little size and temperature dependence.
\par
Another criterion for the critical point $\delta_{\rm c}$ is that the staggered susceptibility $\chi(q=\pi;T)$ is a maximum at the critical point at sufficiently low temperatures\cite{divergeSS}.
We show the $\delta$-dependence of the staggered susceptibility in Fig.\ref{UsSsdelta}(b). Near $\delta\simeq0.2595$, $\chi(q=\pi;T)$ also reaches a maximum with little size and temperature dependence. Thus, we estimate the critical value of $\delta$ as $\delta_{\rm c}=0.2595\pm0.0005$\cite{Kitazawa}.
\par
In order to show the validity of this estimation, we compare the low-temperature behaviors of $\chi(q=0;T)$ and $\chi(q=\pi;T)$ at $\delta=0.2595$ with those at $\delta=0.250$ and $0.255$ in Fig.\ref{UsSslt}. This figure clearly shows that $\delta=0.2595$ is closer to the critical point $\delta_{\rm c}$ than $\delta=0.250$ or $\delta=0.255$.

\section{Low-temperature properties at the critical point}
\label{Main}
In this section, we investigate the low-temperature properties of S1BA at the critical point, assuming that it is described by the $k=1$ SU(2) WZW model (with a marginally irrelevant operator)\cite{EAT}.
\subsection{Uniform susceptibility}
Before investigating the low-temperature behavior of the uniform susceptibility of S1BA at the gapless point, we briefly review the case of the spin-1/2 antiferromagnetic Heisenberg chain (S1/2AH) as an example which is described by the $k=1$ SU(2) WZW model.
Figure \ref{UsS1/2AH} shows the temperature dependence of the uniform susceptibility $\chi(q=0;T)$ for S1/2AH\cite{EAT}. The dotted line corresponds to the fit assuming eq.(\ref{logeq}), which is expected from the renormalization group approach\cite{AGSZ,Nomura,EAT}.
\begin{equation}
\chi(q=0;T)=\frac{1}{2\pi v}+\frac{1}{4\pi v}[\frac{1}{\ln(T_0/T)}-\frac{\ln(\ln(T_0/T)+1/2)}{2[\ln(T_0/T)]^2}]+o(1/[\ln(T_0/T)]^2)
\label{logeq}
\end{equation}
The second term in eq.(\ref{logeq}) is the leading logarithmic correction term due to the marginally irrelevant operator ($\propto\vec{J_L}\cdot\vec{J_R}$).
One of the features due to this logarithmic correction is the infinite slope in the low-temperature limit. As a result, naive extrapolation of $\chi(q=0;T)$ as $T\rightarrow 0$, using the data in the low-temperature regime ($0.02\buildrel < \over \sim T \buildrel < \over \sim 0.2$), does not coincide with the zero-temperature uniform susceptibility $\chi(q=0;T=0)=1/(2\pi v)$\cite{EAT}. Another feature is the existence of an inflection point in the low-temperature regime. For S1/2AH, the inflection point is near $T=0.087$\cite{EAT}.
\par
Let us consider the S1BA case. Figure \ref{UsS1BA} shows the temperature dependence of the uniform susceptibility $\chi(q=0;T)$ for S1BA at the critical point ($\delta=0.2595$). This figure supports the existence of the logarithmic correction in the sense that naive extrapolation of $\chi(q=0;T)$ as $T\rightarrow 0$ does not coincide with the zero-temperature uniform susceptibility $\chi(q=0;T=0)=1/(2\pi v)$, and that there exists an inflection point near $T\simeq0.2$. We have tried to fit the numerical data as eq.(\ref{logeq}), and estimated $T_0\simeq 0.34$. This value of $T_0$ is smaller than that of S1/2AH by about one order of magnitude\cite{EAT}.

\subsection{Staggered susceptibility and staggered structure factor}
Next, we consider the low-temperature behaviors of the staggered susceptibility $\chi(q=\pi;T)$ and the staggered structure factor $S(q=\pi;T)$.
For S1/2AH, Starykh $et.al.$ confirmed that $\chi(q=\pi;T)$ and $S(q=\pi;T)$ in the low-temperature regime behave as follows\cite{Starykh}:
\begin{eqnarray}
\chi(q=\pi;T)&\propto&T^{-1}[\ln(T_{\chi}/T)]^{1/2},\label{ScalingSs}\\
S(q=\pi;T)&\propto&[\ln(T_S/T)]^{3/2}.\label{ScalingS0}
\end{eqnarray}
Here, we consider the S1BA case. Figures \ref{SsS1BA} and \ref{S0S1BA} show the temperature dependence of the staggered susceptibility $\chi(q=\pi;T)$ and the staggered structure factor $S(q=\pi;T)$ for S1BA at the gapless point in the low temperature regime. These figures suggest that the low-temperature behaviors of $\chi(q=\pi;T)$ and $S(q=\pi;T)$ for S1BA at the critical point are qualitatively the same as those of S1/2AH. [For comparison, in the insets of Figs. \ref{SsS1BA}(b) and \ref{S0S1BA}(b), we show the S1/2AH case.]
Thus, the behaviors of $\chi(q=\pi;T)$ and $S(q=\pi;T)$ also support that S1BA at the critical point belongs to the universality class of the $k=1$ SU(2) WZW model.

\section{Comparison with experimental data}
\label{Experiment}
In this section, numerical results are compared with the experimental data for [\{Ni(333-tet)($\mu$-N$_3$)\}$_n$](ClO$_4$)$_n$\cite{HagiwaraUs}. Figure \ref{exp} shows the temperature dependence of the uniform susceptibility for the powder sample of [\{Ni(333-tet)($\mu$-N$_3$)\}$_n$](ClO$_4$)$_n$. 
The global feature of the experimental data is very similar to that of the numerical results for S1BA at the gapless point. This suggests that this compound is effectively described by S1BA near the critical point.
A little difference in the low temperature regime may be due to small anisotropy effects or off-criticality of this compound. 
\section{Summary}
\label{Summary}

We have reported the numerical results for the spin-1 antiferromagnetic Heisenberg chain with bond-alternation (S1BA) obtained by quantum Monte Carlo (loop algorithm). 
We have estimated the strength of bond-alternation at the gapless point as $\delta_{\rm c}=0.2595\pm0.0005$. 
At the gapless point, we have confirmed that the low-temperature properties are effectively described by the $k=1$ SU(2) WZW model (with a marginally irrelevant operator).
The numerical results for S1BA at the critical point well explain recent experimental results for [\{Ni(333-tet)($\mu$-N$_3$)\}$_n$](ClO$_4$)$_n$, indicating that this compound is close to the gapless point of the Affleck-Haldane conjecture. 

\narrowtext
\acknowledgments
The authors would like to thank K.Kawano and K.Nomura for helpful discussions 
and useful comments. One of the authors (M.K.) thanks D. Lidsky for reading of the manuscript.  
Part of the calculations were performed on the Intel Japan PARAGON
at Institute for Solid State Physics, Univ. of Tokyo. 
This research is supported 
in part by Grants-in-Aid for Scientific Research Fund from the Ministry of 
Education, Science and Culture (08640445).

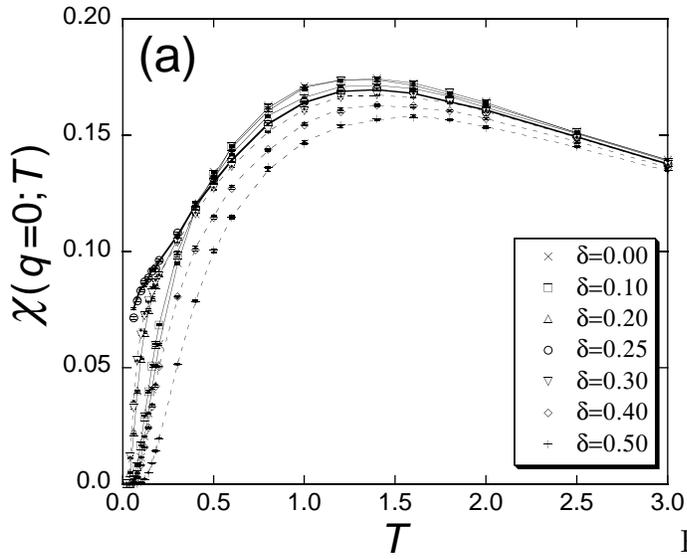
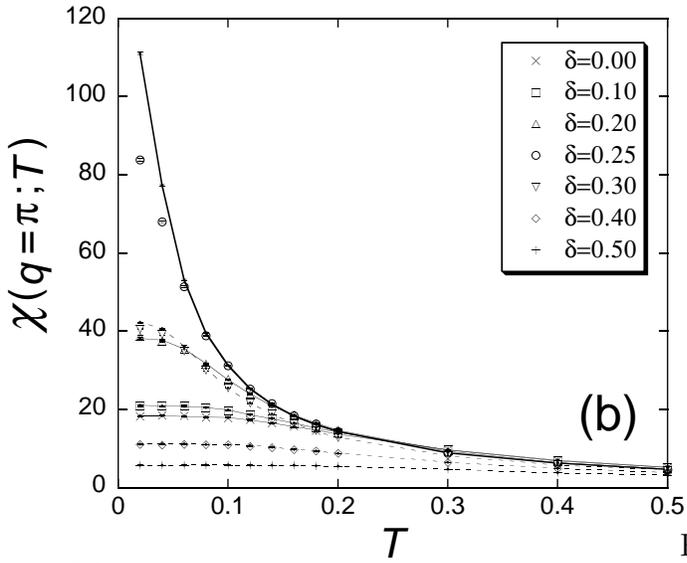
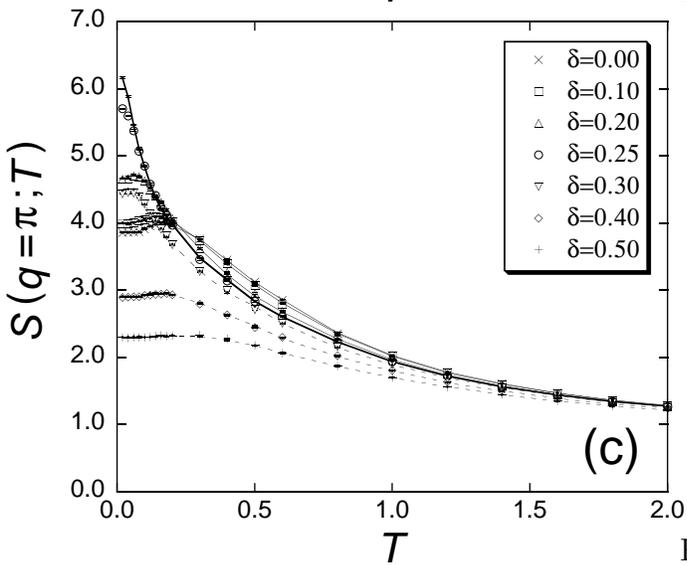
\begin{figure}
\caption{Uniform susceptibility $\chi(q=0;T)$ [(a)], staggered susceptibility $\chi(q=\pi;T)$ [(b)] and the staggered structure factor $S(q=\pi;T)$ [(c)] of S1BA as a function of temperature $T$. 
The symbols are 64-site data. The 96-site date for $0\le\delta<0.25$ ($0.25<\delta\le 0.5$) are joined by solid (dotted) lines. The thick solid line corresponds to 96-site data for $\delta=0.25$.}
\label{UsSsS0gl}
\end{figure}

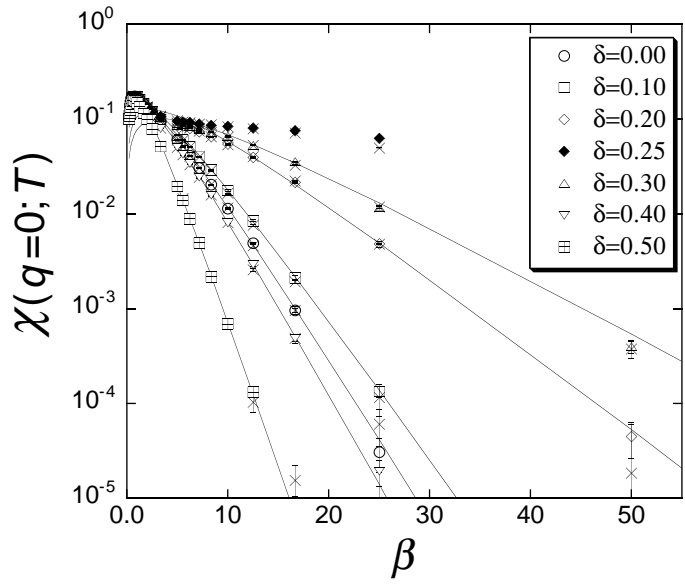
\begin{figure}
\caption{Uniform susceptibility $\chi(q=0;T)$ of S1BA as a function of inverse temperature $\beta$. The solid lines correspond to the fit as $\chi(q=0;T)\propto\frac{1}{\protect\sqrt{T}}\exp(-\Delta/T)$. Open and solid symbols denote 96-site data. Crosses denote 64-site data.}
\label{Usglbeta}
\end{figure}

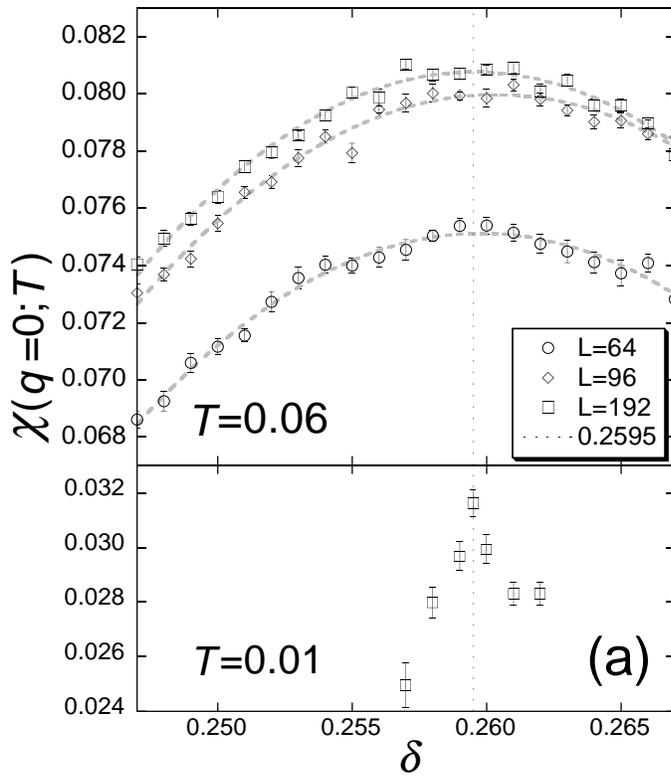
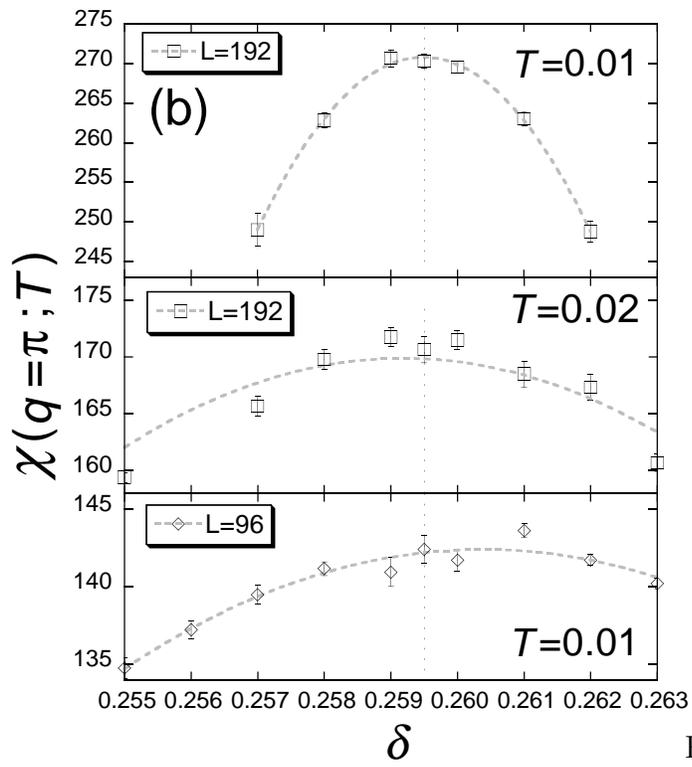
\begin{figure}
\caption{Uniform [(a)] and staggered [(b)] susceptibility of S1BA as a function of $\delta$. The bold lines are guides to the eye.}
\label{UsSsdelta}
\end{figure}

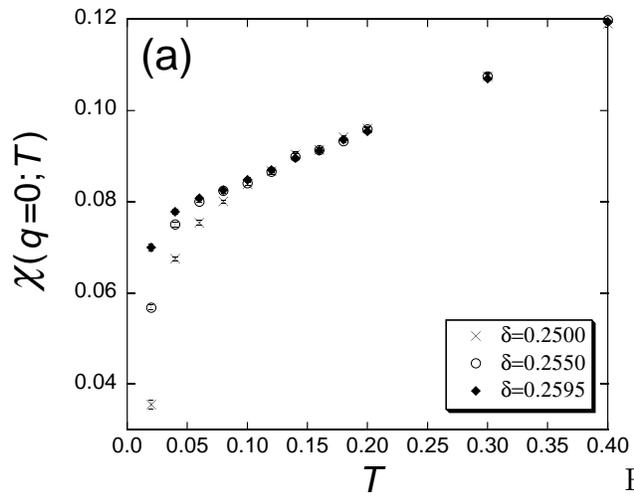
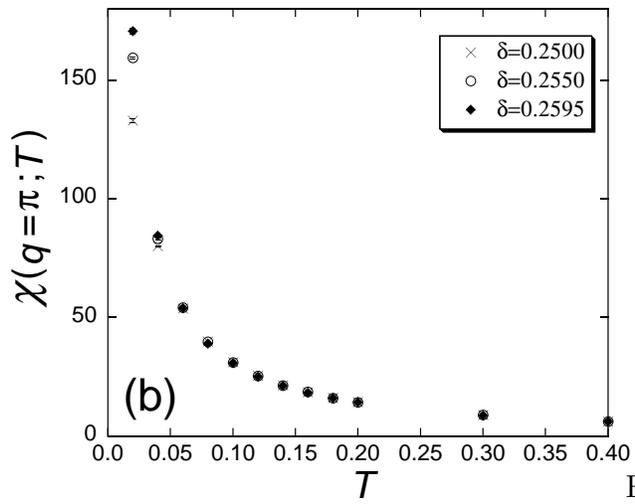
\begin{figure}
\caption{Uniform [(a)] and staggered [(b)] susceptibility of S1BA near the gapless point in the low-temperature regime. $L=192$.}
\label{UsSslt}
\end{figure}

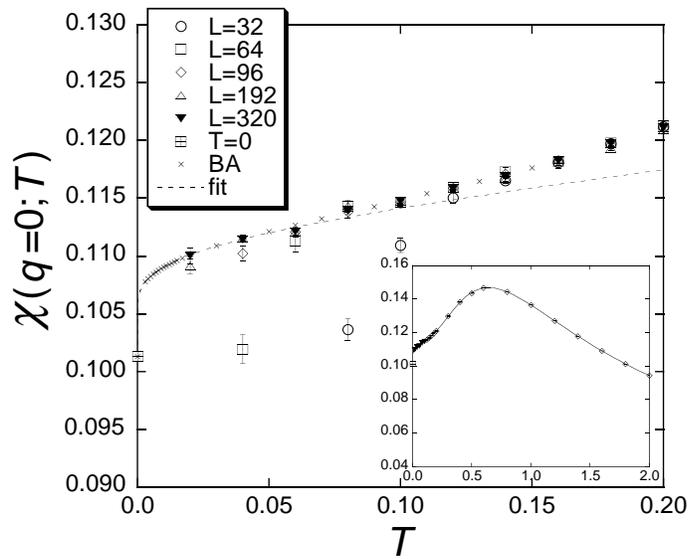
\begin{figure}
\caption{Uniform susceptibility $\chi(q=0;T)$ of S1/2AH for $0\le T \le 0.2$. The crosses denote the result obtained by Bethe Ansatz cited from ref.\protect\cite{EAT}. The dotted line corresponds to the fit assuming eq.(\protect\ref{logeq}). We choose $T_0=2.3$. Solid and open symbols denote the data obtained by QMC loop algorithm. The inset shows $\chi(q=0;T)$ of S1/2AH for $0\le T \le 2$. The solid line in the inset is obtained by Bethe Ansatz\protect\cite{EAT}.}
\label{UsS1/2AH}
\end{figure}

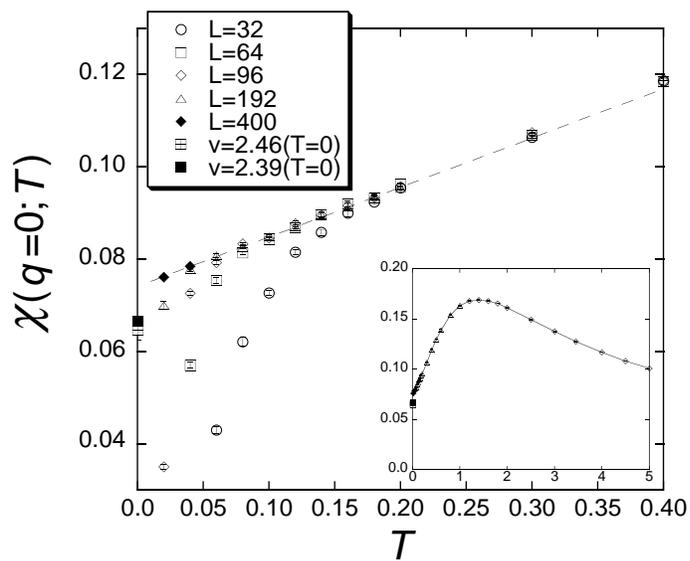
\begin{figure}
\caption{Uniform susceptibility $\chi(q=0;T)$ of S1BA at $\delta=0.2595$ for $0\le T \le 0.4$. The data at $T=0$ are obtained as $\chi(q=0;T=0)=1/(2\pi v)$, where $v=2.46\pm0.08$\protect\cite{Yamamoto1} and $v=2.39$\protect\cite{Totsuka}. The dashed line corresponds to the linear fit using the data for $0.02\le T \le 0.2$. The inset shows $\chi(q=0;T)$ for $0\le T \le 5$.}
\label{UsS1BA}
\end{figure}

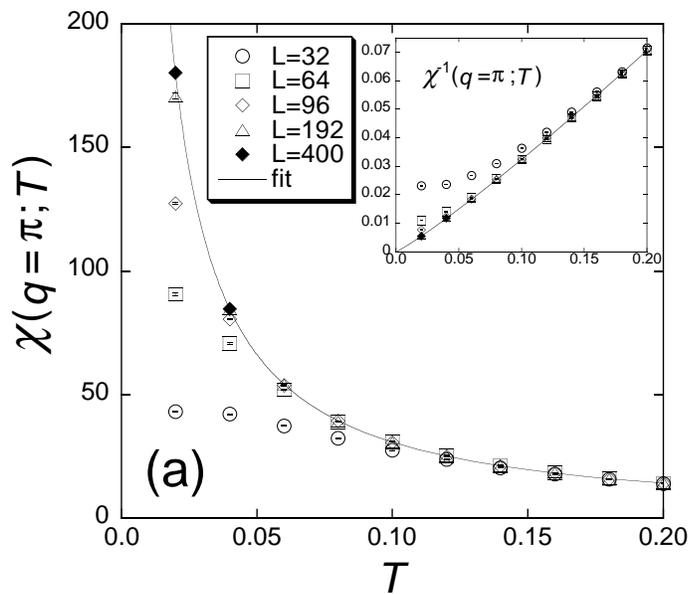
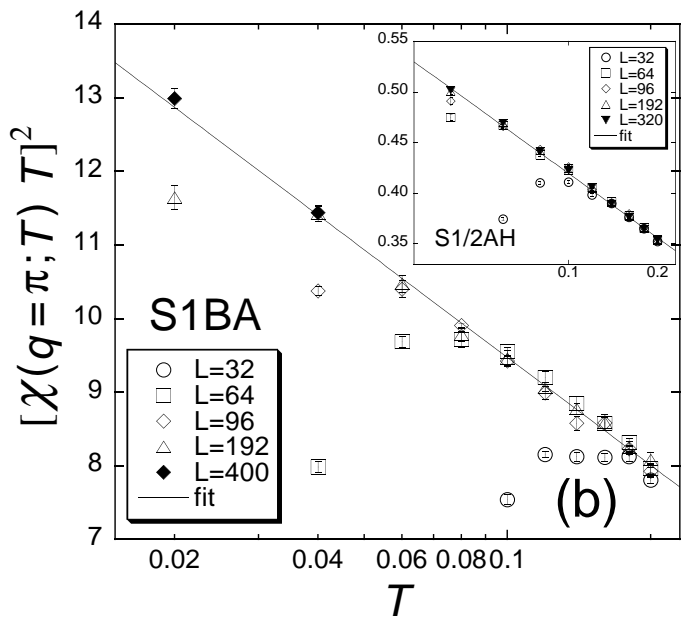
\begin{figure}
\caption{Staggered susceptibility $\chi(q=\pi;T)$ of S1BA at $\delta=0.2595$ in the low-temperature regime. (a) Linear plot and (b) logarithmic plot. The solid line corresponds to the fit assuming eq.(\protect\ref{ScalingSs}). We estimate $T_{\chi}\simeq 8.9$. The inset in (a) shows the inverse of $\chi(q=\pi;T)$. The inset in (b) shows the logarithmic plot of $\chi(q=\pi;T)$ for S1/2AH obtained by QMC loop algorithm. We estimate $T_{\chi}\simeq 9.8$ for S1/2AH.}
\label{SsS1BA}
\end{figure}

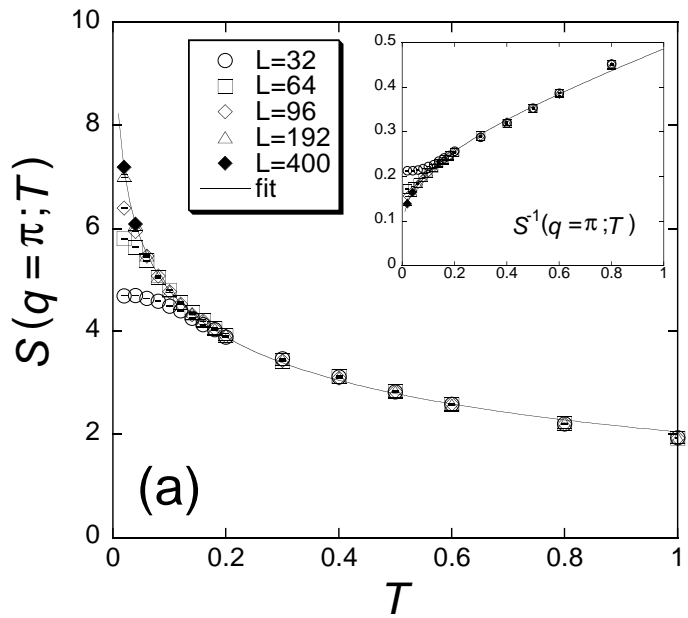
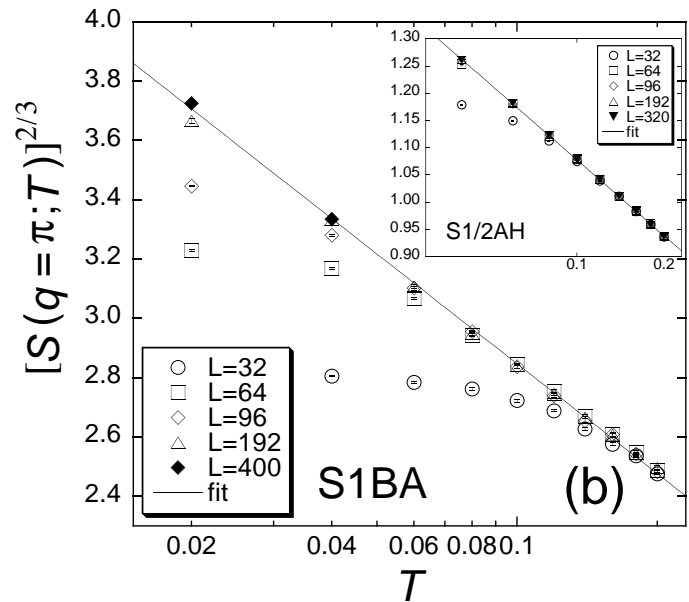
\begin{figure}
\caption{Staggered structure factor $S(q=\pi;T)$ of S1BA at $\delta=0.2595$ in the low-temperature regime. (a) Linear plot and (b) logarithmic plot. The solid line corresponds to the fit assuming eq.(\protect\ref{ScalingS0}). We estimate $T_S\simeq 20.6$. The inset in (a) shows the inverse of $S(q=\pi;T)$. The inset in (b) shows the logarithmic plot of $S(q=\pi;T)$ for S1/2AH obtained by QMC loop algorithm. We estimate $T_S\simeq 21.5$ for S1/2AH.}
\label{S0S1BA}
\end{figure}

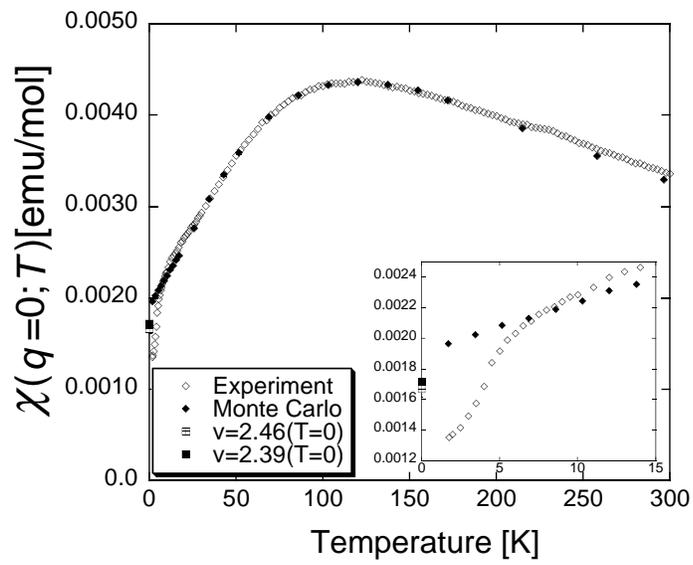
\begin{figure}
\caption{Uniform susceptibility $\chi(q=0;T)$ of [\{Ni(333-tet)($\mu$-N$_3$)\}$_n$](ClO$_4$)$_n$ (open diamonds)\protect\cite{HagiwaraUs} and that of S1BA for $\delta=0.2595$ (solid diamonds). The data at $T=0$ are obtained as $\chi(q=0;T=0)=1/(2\pi v)$, where $v=2.46\pm0.08$\protect\cite{Yamamoto1} and $v=2.39$\protect\cite{Totsuka}. The inset shows the blow-up region for the low-temperature regime. We choose the $g$-value as $g=2.46$, and $J/k_{\rm B}=86$[K]}
\label{exp}
\end{figure}

\newpage
\hspace{-6.2mm}\epsfxsize=3.5in\epsfbox{s1BA1FIG.dir/UsTgl.epsi}
Fig.\ref{UsSsS0gl}(a)
\epsfxsize=3.5in\epsfbox{s1BA1FIG.dir/SsTgl.epsi}
Fig.\ref{UsSsS0gl}(b)
\epsfxsize=3.5in\epsfbox{s1BA1FIG.dir/S0Tgl.epsi}
Fig.\ref{UsSsS0gl}(c)
\epsfxsize=3.5in\epsfbox{s1BA1FIG.dir/Usbgl25.epsi}
\vspace{2mm}\\
Fig.\ref{Usglbeta}\\
\vspace{25mm}\\
\epsfxsize=3.5in\epsfbox{s1BA1FIG.dir/Usdlt.epsi}
Fig.\ref{UsSsdelta}(a)\\
\vspace{2mm}\\
\epsfxsize=3.5in\epsfbox{s1BA1FIG.dir/Ssdlt.epsi}
Fig.\ref{UsSsdelta}(b)\\
\vspace{2mm}\\
\newpage
\hspace{-6.2mm}\epsfxsize=3.2in\epsfbox{s1BA1FIG.dir/UsTnear.epsi}
Fig.\ref{UsSslt}(a)\\
\vspace{2mm}\\
\epsfxsize=3.2in\epsfbox{s1BA1FIG.dir/SsTnear.epsi}
Fig.\ref{UsSslt}(b)\\
\vspace{2mm}\\
\newpage
\hspace{-6.2mm}\epsfxsize=3.5in\epsfbox{s1BA1FIG.dir/UsTS12AH.epsi}
\vspace{2mm}\\
Fig.\ref{UsS1/2AH}\\
\vspace{40mm}\\
\epsfxsize=3.5in\epsfbox{s1BA1FIG.dir/UsTS1BA.epsi}
\vspace{2mm}\\
Fig.\ref{UsS1BA}\\
\vspace{2mm}\\
\epsfxsize=3.5in\epsfbox{s1BA1FIG.dir/SsTS1BA.epsi}
\vspace{2mm}\\
Fig.\ref{SsS1BA}(a)\\
\vspace{35mm}\\
\epsfxsize=3.5in\epsfbox{s1BA1FIG.dir/SsT_red.epsi}
\vspace{2mm}\\
Fig.\ref{SsS1BA}(b)\\
\vspace{2mm}\\
\epsfxsize=3.5in\epsfbox{s1BA1FIG.dir/S0TS1BA.epsi}
\vspace{2mm}\\
Fig.\ref{S0S1BA}(a)\\
\vspace{35mm}\\
\epsfxsize=3.5in\epsfbox{s1BA1FIG.dir/S0T_red.epsi}
\vspace{2mm}\\
Fig.\ref{S0S1BA}(b)\\
\vspace{2mm}\\
\epsfxsize=3.5in\epsfbox{s1BA1FIG.dir/Exp.epsi}
\vspace{2mm}\\
Fig.\ref{exp}


\begin{references}
\bibitem{divergeSS} The staggered susceptibility for the gapless point is expected to diverge at $T=0$ and $L=\infty$.
\label{divergeSS}
\end{references}
\end{document}